# A well-tempered density functional theory of electrons in molecules

Ester Livshits and Roi Baer*

*Department of Physical Chemistry and the Fritz Haber Center for Molecular Dynamics, the Hebrew University of Jerusalem, Jerusalem 91904 Israel.*

Reporting extensions of a recently developed approach to density functional theory with correct long-range behavior (Phys. Rev. Lett. 94, 043002 (2005)). The central quantities are a splitting functional $\gamma[n]$ and a complementary exchange-correlation functional $E_{XC}^{\gamma}[n]$. We give a practical method for determining the value of $\gamma$ in molecules, assuming an approximation for $E_{XC}^{\gamma}$ is given. The resulting theory shows good ability to reproduce the ionization potentials for various molecules. However it is not of sufficient accuracy for forming a satisfactory framework for studying molecular properties. A somewhat different approach is then adopted, which depends on a density-independent $\gamma$ and an additional parameter $w$ eliminating part of the local exchange functional. The values of these two parameters are obtained by best-fitting to experimental atomization energies and bond-lengths of the molecules in the G2(1) database. The optimized values are $\gamma = 0.5 a_0^{-1}$ and $w = 0.1$. We then examine the performance of this slightly semi-empirical functional for a variety of molecular properties, comparing to related works and to experiment. We show that this approach can be used for describing in a satisfactory manner a broad range of molecular properties, be they static or dynamic. Most satisfactory is the ability to describe valence, Rydberg and inter-molecular charge-transfer excitations.

## I. INTRODUCTION

In density functional theory, a system of non-interacting Fermi particles is set up to have the same density as a system of "usual" electrons, interacting through the Coulomb repulsion. The non-interacting particles must be subject not only to the attractive forces of the nuclei of the molecule but also to artificial forces that are designed to ensure the equality of densities. These forces include an average (Hartree) mean-field force and an additional force described by the "exchange-correlation (XC) potential". The XC potential is derived from a universal density functional, the Kohn-Sham exchange-correlation (XC) density functional.[1-4] This functional is extremely complicated and intricate on which barely a handful exact or nearly exact properties and sum rules are known (so called "formal properties"[5]). To build a useful theory based on the density, humans must rely on simple approximations to the XC functional which abide to as many as possible exact properties.

Any practical density functional strives to, but actually cannot ever achieve the Platonic ideal of the Kohn-Sham XC functional. Nevertheless, there has been great success in developing a series of "approximate XC functionals" which grow in complexity but allow more formal properties to be satisfied and offer increasingly greater precision.[5] These start off from local density functionals (LSDA) based on the known properties of the homogeneous electron gas (HEG)[2,6-9], moving into the realm of inhomogeneous densities, both for exchange and correlation, described by the density gradient[2,10-15] (gradient expansion and generalized gradient approximations GEA, GGA, respectively) and then reaching orbital functionals that include explicit forms of Hartree-Fock exchange[4,16] and/or the non-interacting kinetic energy density[17] (often referred to as meta-GGA). Application of these approximations have had great success in describing the chemical bond to good accuracy, including the equilibrium structure, the bond energy and the vibrational-rotational properties of a broad variety of molecules.[5,18-22]

However, with increased testing, it has become evident that whenever quantities other than energy are computed, namely static or dynamical response properties, many popular approximations to the functionals lead to unacceptably large errors or even wrong physical behavior. Examples of such failures are: 1) Static polarizability for example is often greatly exaggerated in elongated molecules;[23] 2) the electronic charge distri-bution does not allocate integer electron charge to weakly interacting subsystems as it should;[24,25] 3) excitation energies to states of charge transfer character, are poorly described[26] 4) Rydberg excitation energies are usually severely underestimated; 5) Small anions are often erroneously predicted to be unstable[27].

These problems in the popular approximate functionals of DFT have been studied in the past and were found related to self-interaction errors[13,28,29]. They hamper many potential uses of density functional theory: studying highly excited electronic states, charge transfer reactions in biological systems and photo-harvesting materials and in even molecular electronics, where DFT based predictions sometimes exhibit huge errors in predicted currents.

To the practical problems mentioned above, we add the related issue, namely the failure to reproduce exactly established formal consequences of the exact functionals. For example, 6) the effective Kohn-Sham potential $v_s(\mathbf{r})$ must, in the large $r$ limit be of the form[30] $const + O(-1/r)$; 7) for an $N$ electron system, the electron removal energy $E_{gs}[N-1] - E_{gs}[N]$ must be equal exactly to $-\varepsilon^{HOMO}[N]$, the energy of the highest occupied molecular orbital (HOMO) in the non-interacting $N$ electron system;[31] 8) and the effective Kohn-Sham potential must have derivative discontinuities[32] (these ensure the integer electron charge in weakly interacting systems). In many density functionals these exact properties were sacrificed in order to have the theory reproduce exactly the homogeneous electron gas ground-state energy which is known to high accuracy using quantum Monte Carlo methods[33].

Recently, several new functionals that tackle the deficiencies described above, associated with the long-range self-repulsion, were published[34-40]. These approaches were based on the long-range corrected (LC) functional of Ikura et al.[34] which was found to overcome many of the above deficiencies. In order to improve the performance of ground-state calculations, a "Coulomb attenuation method" (CAM) was introduced[35]. The best CAM introduces a functional with outstanding ground-state energy predictions but without the exact long-range behavior. The advantages of the correct long-range behavior are however extremely important for producing a more balanced functional, as is evident in the extremely successful applications of the LC functional TDDFT[41] (here referred to as TDLC).

The development of a more balanced functional is thus the purpose of this paper. The only way we could do this is to stick to the correct long-range behavior, even at the price of deviating slightly from the exact HEG limit. We thus try to find a way to improve the LC performance without loosing the property of correct long-range behavior and without introduction of a large number of parameters. Continuing previous theoretical work[39] where we introduced the splitting density functional $\gamma[n]$ and the complimentary exchange-correlation functional $E_{XC}^{\gamma}[n]$, we now attempt to put these concepts to work. We first study the nature of the functional $\gamma[n]$. For this functional we present some exact results for the HEG. We then give a practical, general method to for determining $\gamma[n]$ for any general non-homogeneous case, assuming an approximation to $E_{XC}^{\gamma}$ is known. The resulting theory is applied to several molecules. Despite reasonable success at describing ionization potentials, we conclude the method is not sufficiently accurate for general use, mainly because of our lack of knowledge of the functional $E_{XC}^{\gamma}[n]$. To continue, we abandon the approach of finding the exact $\gamma$ for each system and instead turn to a semi-empirical 2-parameter approach. This functional is based on LYP correlation energy augmented by subtracting a small amount of the local $\gamma$-exchange energy[14] and a combination of exact long-range exchange with a local exchange. The parameters $\gamma$ and $w$ are determined by a fit to 54 molecules of the G2(1) set[42]. The main feature of the resulting functional is its balanced applicability, or its well-temperedness: while admittedly it is not highly accurate for any particular electronic property, it is *reasonably* accurate for a broad range of such properties. All calculations re-ported in this paper were done using a modified version the quantum chemistry code Q-CHEM 2.0 [43]. The necessary modifications were done with a lot of help and support of Dr. Yihan Shao of the Q-CHEM company.

## II. THEORY

In a recent paper[39] the exchange correlation energy of DFT was shown to equal to:

$$E_{XC}[n] = K_X^\gamma[n] + E_{XC}^\gamma[n] \tag{1}$$

Where the explicit exchange is with respect to a Coulomb tail:

$$K_X^\gamma[n] = -\frac{1}{4}\int |P[n](\mathbf{r},\mathbf{r}')|^2 u_\gamma(|\mathbf{r}-\mathbf{r}'|)d^3rd^3r' \tag{2}$$

$P[n](\mathbf{r},\mathbf{r}')$ is the density matrix of non-interacting particles and $u_\gamma(r) = \frac{erf(\gamma r)}{r}$ is the Coulomb tail (a potential having the Coulomb tail but weakened at the origin). The second term is given by:

$$E_{XC}^\gamma[n] = \langle \Psi_{gs} | \hat{Y}_\gamma | \Psi_{gs} \rangle - \frac{1}{2}\int n(\mathbf{r})n(\mathbf{r}')y_\gamma(|\mathbf{r}-\mathbf{r}'|)d^3rd^3r' \tag{3}$$

Where $y_\gamma(r) = \frac{1}{r} - u_\gamma(r)$ is the complementary error-function screened potential, $Y_\gamma = \frac{1}{2}\sum_{i\neq j} y_\gamma(|\mathbf{r}_i - \mathbf{r}_j|)$ and $\Psi_{gs}$ is the ground state of the system.

When applied to the homogeneous electron gas (HEG), one can separate this into two terms, the remaining exchange energy, expressed in terms of local exchange energy $\varepsilon_X^\gamma$ and the correlation energy, expressed in terms of the local quantity $\varepsilon_C$:

$$E_{XC}^\gamma = E_C^\gamma + \int \varepsilon_X^\gamma(k_F(\mathbf{r}))n(\mathbf{r})d^3r \tag{4}$$

Here, the Savin's exchange energy per particle for the complimentary error-function potential[44]:

$$\varepsilon_X^\gamma(k_F) = \frac{k_F}{2\pi} \times \left[q^2\left\{\frac{2\sqrt{\pi}}{q}erf\left(\frac{1}{q}\right) - 1 + (q^2-2)\left(1-e^{\frac{-1}{q^2}}\right)\right\} - \frac{3}{2}\right] \tag{5}$$

where $q = \gamma/k_F$ and $k_F$ is the local Fermi wave-vector at the density. The correlation functional will be written as:

$$E_C^\gamma[n] = \int \tilde{\varepsilon}_C^\gamma(n(\mathbf{r}),|\nabla n(\mathbf{r})|,...)n(\mathbf{r})d^3r \tag{6}$$

Where the correlation energy per particle is the LYP correlation energy[14] somewhat reduced by subtracting from it a small part of the Savin exchange correlation energy per particle:

$$\tilde{\varepsilon}_C^\gamma(n,|\nabla n|,...) = \varepsilon_C^{LYP}(n,|\nabla n|,...) - w\varepsilon_X^\gamma(k_F(n)) \tag{7}$$

For homogenous densities the parameter $w$ is zero when the proper $\gamma$ is used (see below). However, we find in the calculations described below that for non-homogeneous systems a non-zero (but small) value of $w$ considerably improves the overall performance of the functional.

### A. $\gamma$ OF HOMOGENEOUS DENSITIES

The XC energy functional $E_{XC}^\gamma$ in Eq. (3) can be written as an integral involving the XC pair correlation function $g_{XC}(\mathbf{r};\mathbf{r}')$:

$$E_{XC}^\gamma[n] = \frac{1}{2}\int n(\mathbf{r})g_{XC}(\mathbf{r},\mathbf{r}')y_\gamma(\mathbf{r}-\mathbf{r}')n(\mathbf{r}')d^3rd^3r' \tag{8}$$

For the homogeneous gas, the pair correlation depends only on the distance $|\mathbf{r}-\mathbf{r}'|$ and has thus been parameterized to high accuracy.[45] The $\gamma,XC$-energy per particle in the HEG is then:

$$\varepsilon_{XC}^\gamma(n) = \frac{1}{2}n\int g_{XC}(r)y_\gamma(r)d^3r \tag{9}$$

The exact value of $\gamma$ for a homogeneous density $\gamma$ is thus obtained by requiring:

$$\varepsilon_{XC}^{\gamma}(n) - \varepsilon_{X}^{\gamma}(k_F(n)) = \varepsilon_{LDA}(n) \quad (10)$$

This equation can be solved implicitly for $\gamma(n)$ and the of such a calculation is given in Figure 1. This exact results sheds light on the "mysterious parameter. First, we see that it is strongly density dependent. In metallic densities ($r_s$ usually lies in the range $1 < r_s < 5$) $\gamma$ takes values from the interval $[0.1,1]$. Also, we note that the function is smooth and monotonic. The value oif $\gamma$ for unpolarized case (total spin zero) is higher, but not by much, than the value for the fully spin-polarized gas.

## B. $\gamma$ OF INHOMOGENEOUS DENSITIES

What about inhomogeneous densities like in molecules? It is nearly impossible to compute $E_{XC}^{\gamma}[n]$ for a given non-homogeneous density, so it is tempting to try and use the results for the HEG. But in trying to do so, we stumble on a difficulty of assigning $\gamma$ to the density inhomogeneous $n$. An attempt is to use an "average density" or average $r_s$ met with difficulties as described in detail in ref.[25] We have discovered a practical approach to address this issue as follows. We make use of an exact relation[46]:

$$-\varepsilon^{HOMO}(N) = E_{gs}(N) - E_{gs}(N-1), \quad (11)$$

Then, using an approximate functional for inhomogeneous densities $E_{XC}^{\gamma}$, we search for $\gamma$ that obeys:

$$\Delta E(\gamma) \equiv \varepsilon^{HOMO}(N;\gamma) + [E_{gs}(N;\gamma) - E_{gs}(N-1;\gamma)] = 0 \quad (12)$$

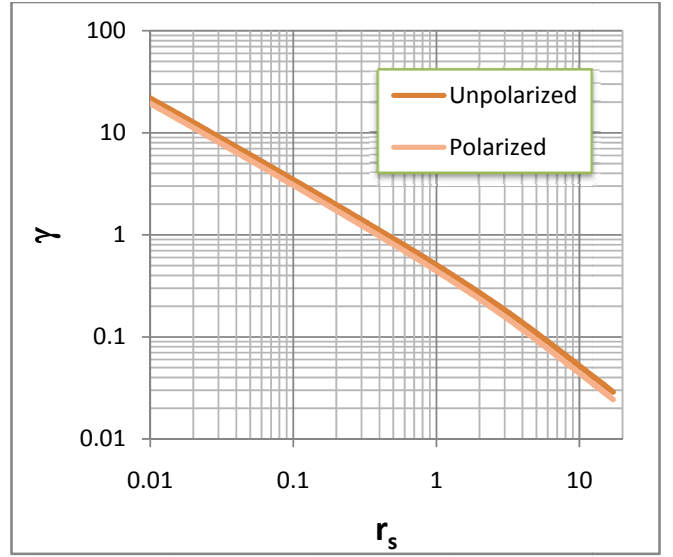

FIGURE 1: THE PARAMETER γ AS A FUNCTION OF THE DENSITY PARAMTER $R_s$ FOR THE FULLY POLARIZED AND UNPOLARIZED HOMOGENEOUS ELECTRON GAS

Besides the use of an approximate XC functional here there is an additional approximation made, since we use in Eq. (12) the same $\gamma$ for calculating both $E[N]$ and $E[N-1]$.

The value of $\gamma$ for which $\Delta E$ is zero was calculated by solving Eq. (12) for a several molecules. As an example of such a calculation, we show in Figure 2 the results of the deviance $\Delta E(\gamma)$ as a function of $\gamma$ for $N_2$ and its close ions. We should note that for the many systems we applied this approach we have found only one case where the $\Delta E(\gamma)$ had no root. This was the cation $Na_2^+$.

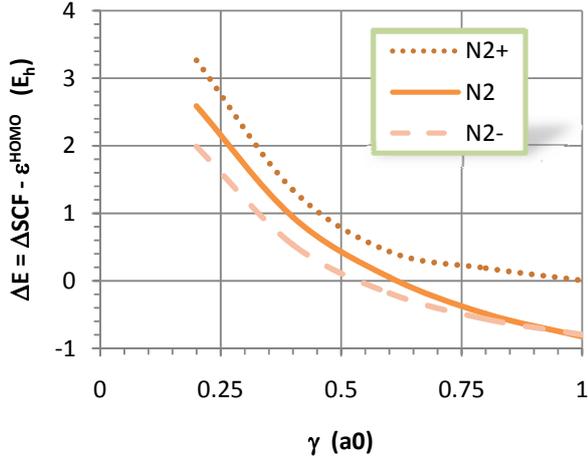

**FIGURE 2: THE DEVIATION** $\Delta E$ **AS A FUNCTION OF** $\gamma$ **(SEE EQ.(12)), FOR N$_2$ AND ITS IONS (CALCULATED USING CC-PVTZ BASIS SET.**

Using the scheme given above, we estimated the values of $\gamma$ for several molecules, as shown in Table 1. We find $\gamma$ in the same range as those of the the homogeneous gas taken at typical valence densities of molecules. One noticeable fact is that the dimers of alkali metal atoms have a low value of $\gamma$. This is not unreasonable since these molecules have low valence electron densities and according to the result for the HEG (Figure 1), lower densities are associated with smaller values of $\gamma$.

To check whether the value of $\gamma$ in Table 1 can lead to a good description of electronic structure of atoms and molecules, we compared the value of the ionization potential (IP) obtained at the canonical $\gamma$ against experimental IP's. This too is shown in Table 1. The calculated vertical IP's agree well with experimental results, however there are some large deviations exaggerating the IP by as much as 1.2eV, occurring at N$_2$, O$_2$ and F$_2$. For other species the deviance is usually considerably smaller. The root-mean-square (RMS) deviance fro experiment is 0.6eV. In an attempt to lower the RMS error, we allowed $w$ in Eq. (7) to assume a non-zero value, namely $w = 0.1$, for which we found significant improvement for the IP that have the same quality as those of B3LYP. Similar quality can be seen in related functionals[35,38,47].

**TABLE 1: AN ESTIMATE OF** $\gamma$ **FOR VARIOUS MOLECULES, OBTAINED BY SATISFYING EQ. (12). THE CALCULATED IP FOR TWO VALUES OF** $w$ **IN EQ. (4) AND THE IP CALCULATED BY B3LYP ARE SHOWN IN COMPARISON WITH EXPERIMENT (ALL ENERGIES IN EV). BASIS SET USED CC-PVTZ.**

| Molecule | $\gamma$ ($a_0^{-1}$) | IP experimental[48,49] | w=0 $-\varepsilon^{HOMO}$ | w=0.1 $-\varepsilon^{HOMO}$ | B3LYP $\Delta SCF$ |
|---|---|---|---|---|---|
| Li$_2$ | 0.3 | 5.1 | -0.1 | 0.0 | -0.2 |
| N$_2$ | 0.6 | 15.6 | -1.0 | -0.6 | 0.0 |
| O$_2$ | 0.7 | 12.1 | -1.2 | -0.9 | -0.7 |
| F$_2$ | 0.7 | 15.7 | -0.8 | -0.4 | 0.1 |
| Na$_2$ | 0.4 | 4.9 | -0.2 | -0.1 | -0.3 |
| P$_2$ [a)] | 0.8 | 10.6 | -0.4 | -0.3 | 0.0 |
| S$_2$ | 0.5 | 9.4 | -0.4 | -0.2 | -0.2 |
| Cl$_2$ | 0.5 | 11.5 | -0.3 | 0.0 | 0.1 |
| BeH [b)] | 0.6 | 8.2 | -0.2 | -0.1 | -0.2 |
| CO | 0.6 | 14.0 | -0.3 | -0.1 | 0.0 |
| HF | 0.7 | 16.0 | -0.4 | 0.0 | 0.0 |
| HCl | 0.5 | 12.8 | 0.0 | 0.3 | 0.2 |
| NH$_3$ | 0.5 | 10.2 | -0.6 | -0.4 | -0.5 |
| PH$_3$ | 0.4 | 9.9 | -0.6 | -0.4 | -0.6 |
| **Mean** | | | **-0.5** | **-0.2** | **-0.2** |
| **RMS** | | | **0.6** | **0.4** | **0.3** |

a) For P$_2$, $\gamma$ changes from 0.8 to 1 $a_0^{-1}$ when the value of $w$ changes from 1 to 0.9.
b) For BeH, $\gamma$ changes from 0.6 to 0.7 $a_0^{-1}$ when the value of $w$ changes from 1 to 0.9.

## C. SIZE CONSISTENCY PROBLEM

The next test of this approach would be to compute atomization energies. Here we immediately stumble upon a problem. Consider an atom $A$ with ground-state density $n_A(\mathbf{r})$ denote $\gamma_A = \gamma[n_A]$ and an atom $B$ with density $n_B(\mathbf{r})$ and $\gamma_B = \gamma[n_B]$. When both atoms are taken together at infinite separation the density is $n_{A\cdots B}(\mathbf{r}) = n_A(\mathbf{r}) + n_B(\mathbf{R}+\mathbf{r})$, where $\mathbf{R}$ is very large. Evidently, this system too has a parameter $\gamma_{A\cdots B}$. In general, when we use an approximate functional $E_{XC}^\gamma$, there is no practical way to ensure that size consistency is obeyed:

$$E_{XC}^{\gamma_{A\cdots B}}[n_{A\cdots B}] = E_{XC}^{\gamma_A}[n_A] + E_{XC}^{\gamma_B}[n_B] \qquad (13)$$

This is in contrast to the case when $\gamma$ is independent of the density. Breaking size consistency is usually unacceptable and it has serious consequences for atomization energy calculations.

## III. A SEMI-EMPIRICAL FUNCTIONAL

The size consistency problem discussed in the previous section and also the insufficiently accurate results for atomization energies prompt us to search for a single value of $\gamma$ applicable for all systems. Since we do not have the exact form of $E_{XC}^\gamma$ this is a necessity. Based on the improvement for the IP we saw in the previous section when we subtracted from the correlation energy a small part of the local exchange energy, (using a factor of w = 0.1) we now try a 2-parameter functional given by $E_{XC}^\gamma$ of Eqs. (4)-(7) were $w$ and $\gamma$ are constants determined by fitting to basic thermochemical (experimental) data in a small but encompassing list of molecules, such as G2(1), including 54 molecules[42]. We performed the following procedure, inspired by other semi-empirical functionals[4]. For each molecule in the G2(1) list we computed the minimizing nuclear configuration (bond lengths) $R_i(\gamma,w)$ and the corresponding atomization energy $D_i(\gamma,w)$, where $i=1,2,...,54$. This was done for several values of $\gamma$ and $w$. The quality of correspondence to experiment is obtained from the RMS errors (L2 norm):

$$\Delta_D(\gamma,\omega)^2 = \frac{1}{54}\sum_{i=1}^{54}\left(D_i(\gamma,\omega) - D_i(\exp)\right)^2$$
$$\Delta_R(\gamma,\omega)^2 = \frac{1}{54}\sum_{i=1}^{54}\left(R_i(\gamma,\omega) - R_i(\exp)\right)^2 \quad (14)$$

A single number summarizing this is the average of relative errors:

$$\Delta(\gamma,w) = \frac{1}{2}\left(\frac{\Delta_D}{\min_i D_i} + \frac{\Delta_R}{\min_i R_i}\right) \quad (15)$$

The minimizing combination was found to be:

$$\gamma = 0.5 a_0^{-1} \qquad w = 0.1 \quad (16)$$

**TABLE 2: PERFORMANCE ON G2(1) DATABASE. CC-PVTZ BASIS WAS USED.**

|  | Atomization energy | | | Bond length | | |
|---|---|---|---|---|---|---|
|  | Exp[a] | This | B3LYP | Exp[b] | This | B3LYP |
| $H_2$ | -104 | 1% | 6% | 0.741 | 4% | 0% |
| LiH | -58 | -1% | 0% | 1.596 | 1% | 0% |
| CN | -179 | -3% | -1% | 1.172 | -1% | 0% |
| $N_2$ | -227 | 2% | 0% | 1.098 | 0% | 0% |
| $F_2$ | -38 | 23% | -1% | 1.412 | -2% | -1% |
| $Cl_2$ | -57 | -10% | -7% | 1.987 | 1% | 2% |
| $O_2$ | -118 | 13% | 4% | 1.208 | -1% | 0% |
| $P_2$ | -116 | -10% | -2% | 1.893 | -1% | 0% |
| $S_2$ | -98 | -5% | 2% | 1.889 | 0% | 2% |
| $Si_2$ | -74 | -10% | -1% | 2.246 | 0% | 2% |
| BeH | -48 | 16% | 18% | 1.343 | 2% | 0% |
| $Li_2$ | -26 | -21% | -20% | 2.673 | 2% | 1% |
| CH | -84 | -4% | 1% | 1.120 | 2% | 1% |
| $^3CH_2$ | -189 | -1% | 0% | 1.085 | 2% | 0% |
| $CH_3$ | -306 | -1% | 0% | 1.079 | 2% | 0% |
| $CH_4$ | -420 | -2% | -1% | 1.094 | 2% | 0% |
| NH | -82 | 2% | 6% | 1.036 | 3% | 1% |
| $NH_2$ | -182 | 0% | 2% | 1.024 | 2% | 1% |
| $NH_3$ | -297 | -1% | 0% | 1.012 | 2% | 1% |
| OH | -107 | 0% | -1% | 0.970 | 3% | 1% |
| $H_2O$ | -233 | -2% | -3% | 0.958 | 2% | 1% |
| HF | -142 | -2% | -4% | 0.917 | 3% | 1% |
| LiF | -139 | 2% | -4% | 1.564 | 1% | 1% |
| $C_2H_2$ | -404 | -2% | -1% | 1.203 | 0% | 0% |
| $C_2H_4$ | -562 | -2% | -1% | 1.339 | -1% | -1% |
| $C_2H_6$ | -711 | -1% | -1% | 1.536 | 0% | 0% |
| HCN | -313 | -1% | -1% | 1.064 | 3% | 1% |
| CO | -261 | 0% | -3% | 1.128 | 0% | 0% |
| HCO | -279 | 1% | -1% | 1.080 | 6% | 5% |
| $H_2CO$ | -376 | -1% | -2% | 1.111 | 2% | 0% |
| $H_3COH$ | -513 | -1% | -2% | 0.956 | 3% | 1% |
| NO | -153 | 6% | 1% | 1.151 | 0% | 0% |
| $N_2H_4$ | -437 | 1% | 0% | 1.446 | -1% | 0% |
| $H_2O_2$ | -268 | 2% | -2% | 1.475 | -3% | -1% |
| $CO_2$ | -392 | 1% | -2% | 1.162 | 1% | 0% |
| $SiH_2$ | -154 | -3% | -1% | 1.514 | 2% | 1% |
| $SiH_3$ | -226 | -1% | 0% | 1.468 | 3% | 2% |
| $SiH_4$ | -324 | -2% | -1% | 1.480 | 2% | 1% |
| $PH_2$ | -153 | 1% | 2% | 1.428 | 1% | 0% |
| $PH_3$ | -241 | -1% | 0% | 1.421 | 1% | 0% |
| $SH_2$ | -182 | -3% | -1% | 1.328 | 3% | 2% |
| HCl | -107 | -4% | -3% | 1.275 | 2% | 1% |
| $Na_2$ | -19 | -20% | -12% | 3.079 | -2% | -1% |
| NaCl | -99 | -8% | -9% | 2.361 | 1% | 1% |
| SO | -122 | 2% | 0% | 1.481 | 1% | 2% |
| SiO | -191 | -2% | -4% | 1.510 | 1% | 1% |
| CS | -172 | -9% | -5% | 1.535 | 0% | 1% |
| ClF | -62 | 2% | -7% | 1.628 | 0% | 2% |
| ClO | -62 | 2% | 1% | 1.570 | 1% | 2% |
| $Si_2H_6$ | -533 | -2% | -2% | 1.470 | 3% | 1% |
| $H_3CSH$ | -473 | 2% | -1% | 1.329 | 1% | 2% |
| $H_3CCl$ | -395 | -2% | -2% | 1.785 | 0% | 1% |
| **Mean** |  | -1% | -1% |  | 1% | 0.7% |
| **RMS** |  | 7% | 5% |  | 2% | 1% |
| (a) kcal/mole | (b) Å | | | | | |

Using other norms than L2, i.e. L1 norm, max norm and relative errors gave similar results although the optimal value of $\gamma$ and $w$ changed a bit, the choice of Eq. (16)

always lead to almost identical performance. We thus chose the parameter values of Eq. (16) and these will be used throughout the rest of the paper. The resulting functional will be referred to in this paper as "This" functional. The performance of This functional for atomization energies and bond lengths is shown in Table 2. Since the atomization energy is proportional to the number of bonds in the molecule and since bond lengths in different molecules change within a factor of 3 throughout the database, we present here the relative errors. Comparing This functional with B3LYP, we see that the latter is clearly superior in most cases. The RMS error in bond lengths is 2% for This functional, considerably larger than in B3LYP (1%) but is still very useful. The RMS atomization energy of This functional is 7% while that of B3lYP is 5%. Next, consider harmonic vibrational wavenumbers for several diatomic molecules shown in Table 3. We compare results derived from This functional to B3LYP and experiment. The performance of This functional is much less satisfying than B3LYP. Evidently, there is a tendency to overestimate the frequencies, by 6% on the average. The RMS error in This functional is much larger than B3LYP and reaches 10%. CAM-B3LYP description of vibrational frequencies was recently examined,[36] showing a much smaller RMS error than This functional although it too has a strong tendency to overestimate the frequencies.

TABLE 3: HARMONIC VIBRATIONAL ENERGIES (CM$^{-1}$) AND THE RELATIVE ERRORS FOR THIS AND B3LYP FUNCTIONALS

|      | *Exp* | *This* | *B3LYP* |
|------|-------|--------|---------|
| H$_2$   | 4401  | -3%    | -1%     |
| LiH  | 1406  | 4%     | 0%      |
| N$_2$   | 2359  | 6%     | 3%      |
| F$_2$   | 917   | 30%    | 10%     |
| Cl$_2$  | 560   | 7%     | -6%     |
| P$_2$   | 781   | 10%    | 2%      |
| Li$_2$  | 351   | 1%     | -3%     |
| HF   | 4138  | 0%     | -1%     |
| LiF  | 910   | 1%     | -1%     |
| CO   | 2170  | 5%     | 1%      |
| **Mean** |   | **6%** | **0.7%** |
| **RMS**  |   | **10%** | **5%** |

We conclude that This functional is obviously inferior to B3LYP, CAM-B3LYP and other related functionals[35,38,47] for describing the chemical bond (energy, and shape of potentials surfaces). However, the results do show that This functional is in fact capable of yielding a reasonably good description of the chemistry. Since we aim at a well-tempered approach, we continue to show that This functional maintains a reasonable description of molecular properties often better than the other functionals.

TABLE 4: IONIZATION POTENTIALS (EV) OF SELECTED SPECIES AND THE DEVIANCE OF THIS AND B3LYP FUNCTIONALS.. BASIS SET: CC-PVTZ.

| Molecule | IP exp[49] | ΔThis $-\varepsilon^{HOMO}$ | ΔB3LYP $-\varepsilon^{HOMO}$ | ΔThis $\Delta SCF$ | ΔB3LYP $\Delta SCF$ |
|---|---|---|---|---|---|
| BeH | 8.2 | -0.1 | -2.8 | 0.1 | 0.2 |
| CH | 10.6 | -0.9 | -3.7 | -1.3 | 0.2 |
| NH | 13.5 | -0.9 | -4.5 | -0.1 | 0.0 |
| OH | 13.0 | -0.9 | -4.5 | -0.1 | 0.0 |
| CN | 13.6 | 0.3 | -3.2 | 0.4 | 0.5 |
| NO | 9.3 | 0.0 | -3.4 | 0.5 | 0.8 |
| F2 | 15.7 | -1.2 | -4.7 | -0.2 | -0.1 |
| Li2 | 5.1 | 0.2 | -1.5 | 0.0 | 0.2 |
| O2 | 12.3 | -0.5 | -3.9 | 0.2 | 0.5 |
| H2O | 12.6 | -0.7 | -4.4 | -0.2 | -0.2 |
| C2H4 | 10.7 | 0.0 | -3.2 | -0.3 | -0.2 |
| H | 13.6 | -1.7 | -5.0 | -0.5 | -0.1 |
| He | 24.4 | -3.3 | -42.0 | -0.5 | 0.4 |
| Li | 5.4 | -0.1 | -1.8 | 0.1 | 0.2 |
| B | 8.3 | -0.2 | -3.3 | 0.2 | 0.3 |
| Be | 9.3 | -0.6 | -3.1 | -0.4 | -0.3 |
| C | 11.3 | -0.7 | -4.2 | 0.0 | 0.1 |
| N | 14.5 | -1.4 | -5.1 | -0.2 | -0.1 |
| O | 13.6 | -1.1 | -4.7 | 0.1 | 0.3 |
| F | 17.4 | -1.9 | -5.6 | 0.0 | 0.1 |
| **Mean** | | **-0.8** | **-5.7** | **-0.1** | **0.1** |
| **RMS** | | **1.0** | **10.0** | **0.4** | **0.3** |

## D. IONIZATION POTENTIALS AND ELECTRON AFFINITY

Equipped with This functional, with 2 parameters that were determined by considering only atomization energies and equilibrium bond lengths, we proceed to examine its predictive power with respect to other properties. First, check the issue of ionization potentials. The results of IP's for the molecules considered also in Table 1 are shown in Table 4. We find that the IP calculated by the $\Delta SCF$ method in when comparison to experiment an overall similar performance as B3LYP.

However, for calculating molecular properties, the closeness of the HOMO energy to the IP is more important. Here This functional is clearly superior to B3LYP, having an RMS smaller by a factor of 10.

A related test is electron affinity (EA). Here, the methods which give up correct long range behavior, such as B3LYP do not do as well. Once again the EA can be estimated in two ways: as the energy difference between the anion and the neutral ($\Delta SCF$) and as the energy of the HOMO orbital of the anion, calculated in the stable ionic nuclear configuration. In Table 5 we show the electron affinity calculated in these two ways. In This functional, the RMS errors in EA are similar in magnitude to the errors in IP. The performance is better than in B3LYP. Like in the case of IP's, the HOMO in B3LYP is a very bad estimate (usually leading to unstable anion prediction in this database). For This, both estimates are reasonable, although the $\Delta SCF$ method has smaller RMS error.

TABLE 5: ELECTRON AFFINITY (EV) OF SELECTED SPECIES AND THE DEVIANCE OF THIS AND B3LYP FUNCTIONALS. BASIS SET: AUG-CC-PVTZ.

| Molecule | IP experimental[49] | $\Delta$This $-\varepsilon^{HOMO}$ | $\Delta$B3LYP $-\varepsilon^{HOMO}$ | $\Delta$This $\Delta SCF$ | $\Delta$B3LYP $\Delta SCF$ |
|---|---|---|---|---|---|
| BeH | 0.7 | 0.1 | -1.7 | -0.4 | -0.2 |
| CH | 1.2 | 0.4 | -2.5 | -0.1 | 0.0 |
| NH | 0.4 | 0.6 | -2.4 | 0.0 | 0.1 |
| OH | 1.8 | 0.5 | -2.9 | 0.0 | -0.1 |
| CN | 3.9 | 0.7 | -2.7 | 0.3 | 0.1 |
| NO | 0.0 | -0.1 | -2.3 | -0.3 | -0.2 |
| $O_2$ | 0.5 | -0.3 | -3.2 | -0.6 | -0.5 |
| H | 0.8 | 0.7 | -1.8 | 0.0 | 0.1 |
| $F_2$ | 3.0 | 2.5 | -1.0 | -0.8 | 3.0 |
| Li | 0.6 | 0.0 | -1.2 | -0.2 | -0.1 |
| B | 0.3 | 0.4 | -1.7 | -0.1 | 0.1 |
| C | 1.3 | 0.4 | -2.5 | -0.1 | 0.0 |
| O | 1.5 | 0.5 | -2.9 | 0.2 | 0.1 |
| F | 3.4 | 0.2 | -3.6 | 0.2 | 0.0 |
| **Mean** | | 0.5 | -2.3 | -0.1 | 0.2 |
| **RMS** | | 0.8 | 2.4 | 0.3 | 0.8 |

E. HYDROGEN BONDING

An important class of chemical bonds with importance to biology and materials science is hydrogen bonds. These involve a relative weak interaction between a hydrogen (which is chemically bonded to an electronegative "donor" atom) and an electronegative "acceptor" atom. In this paper we do not consider the hydrogen bond in great detail and following ref.[36] we mainly test the donor-acceptor distance for a few hydrogen-bond system. We used for this the calculation daug-ccpVTZ basis and the results are shown in Table 6. We find that This functional predicts shorter bond lengths than actually measured (oppositely to B3LYP), similar to CAM-B3LYP[36] with errors on the order of 0.02A. These errors are relatively small (less than 1%) and thus quiet acceptable.

TABLE 6: HYDROGEN BOND DIMER DISTANCES (A) FOR SEVERAL SPECIES

| | ab initio[a)] | This | B3LYP |
|---|---|---|---|
| $(HF)_2$ | 2.73 | -0.03 | 0.02 |
| (CO)(HF) | 2.95 | -0.02 | 0.06 |
| (OC)(HF) | 2.99 | -0.01 | 0.00 |
| $(H_2O)_2$ | 2.93[b)] | -0.03 | 0.01 |
| **Mean** | | -0.02 | 0.02 |
| **RMS** | | 0.02 | 0.03 |

a) Values taken from ref.[36]
b) Smaller ab initio estimates, of 2.91A have also been published[50].

For the water dimer, we also calculated the equilibrium dissociation energy using This functional and obtained 5.9kcal/mole. Comparing to experiment[51] and high-level ab-initio calculations[50,52] our result is about 15-20% too high. Unfortunately the corresponding values calculated using other functionals (LC, CAM-B3LYP) were not reported.

A. STATIC POLARIZABILITY

One of the most important detrimental effects of self repulsion are seen when static polarizability of elongated molecules is estimated by DFT methods. It was found that in this case local functionals lead to greatly exaggerated values[53]. The work of van Faassen et al[54] established benchmark systems for this, examining long linear arrays of various types of oligomer units. One such case is where the oligomer unit is the $H_2$ molecule and another is polyacetylene where each oligomer is $C_2H_4$. We examined these systems using the geometry of ref [54] and a cc-PVDZ basis set. In both cases, it was shown that HFT somewhat overestimates polarizability per unit but gives reasonable results [53,55]. The fact that long-range self repulsion was involved was discussed in [54] and later by others[39,56,57]. In Figure 3 we show our results for these two systems. It is seen that B3LYP suffers from similar ailments as seen in local functionals, although to a smaller degree. In contrast This functional gives results which are close to HFT and are therefore physically correct. Similar quality of results (but on a conjugated system) were reported for the LC functional of Ikura et al[34].

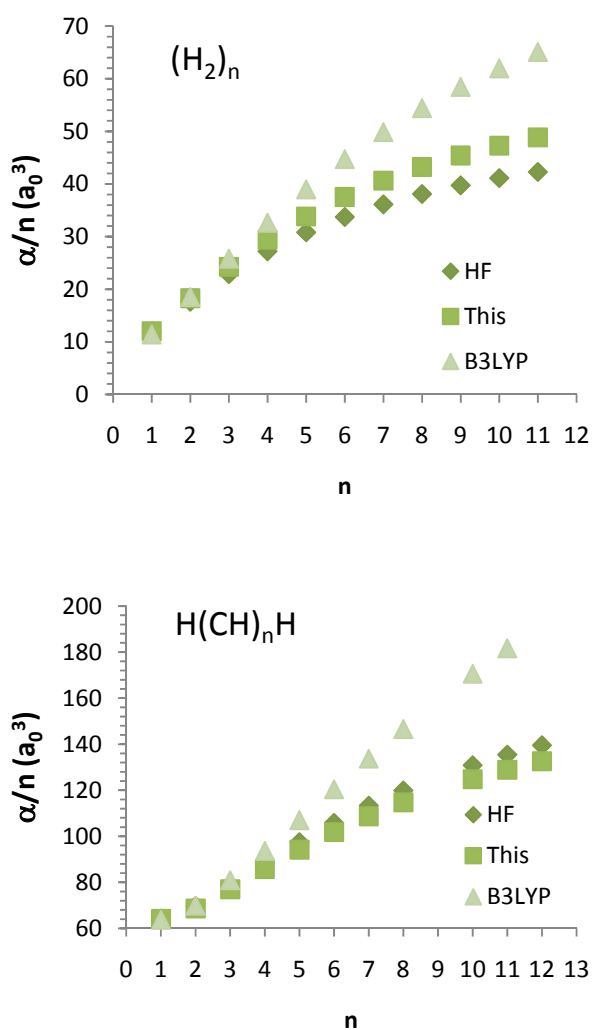

FIGURE 3: POLARIZABILITY ($a_0^3$) AS A FUNCTION OF LENGTH FOR A CHAIN OF $H_2$ MOLECULES (TOP) AND FOR POLYACETYLENE (BOT).

## B. VALENCE AND RYDBERG EXCITATIONS

The use of the DFT functionals as adiabatic functionals for TDDFT is common practice[58-68] (although recently some attempts were made to go beyond the adiabatic functionals[69,70]). For valence excitations having a dominant 1-electron excitation character the results are often of high quality. This will be also seen in the several examples we show here. However, functionals which suffer from long range self-repulsion (LSDA, GGA, B3LYP) will not correctly describe charge-transfer and Rydberg excitations.

To check the performance of This functional for describing excitation energies, we have applied it to several standard benchmark molecules, comparing to experimental excitation energies, to results of TDHF, to adiabatic TDB3LYP and to results reported by Tawada et al [41] (TDLC). The deviance of the different methods from experimental results are shown in Figure 4. All the results, except those of Tawada et al were calculated using the modified version of Q-CHEM 2.0 discussed above.[43] We used a high quality daug-ccpVTZ basis set, except for $C_6H_6$ where the basis set aug-ccpVTZ supplemented with an additional diffuse set centered on nuclear the center of mass, as described in ref [71]. The TDLC results pertain to a Sadlej+ basis[72], which is smaller than ours, We also performed calculations in the Sadlej+ basis set and obtained similar, but somewhat less satisfactory results than shown for the larger more complete basis-set.

We now describe briefly the results shown in Figure 4. As seen, usually TDHF gives very poor results (except for $C_2H_4$). The TDB3LYP functional allows good description of the valence states in some molecules but always severely underestimates Rydberg excitations. Overall, the two theories which correct for long-range repulsion (TDLC and This functionals) seem to be much more balanced than TDB3LYP and TDHF: they allow uniformly reasonably good description of valence and Rydberg excitations. The present functional seems to slightly outperform the TDLC functional in most molecules, except $C_6H_6$. Yet in the latter case, when Rydberg states are compared, a Sadlej+ basis is inappropriate, as center-of-mass orbitals are actually very important[73]. A curious result, which we do not understand, is that This functional also has a large error for the first triplet state of $C_6H_6$. Although for other valence states of this molecules it performs well.

Considering that the parameters of This functional were optimized for the ground state of the G2(1) database, they perform surprisingly well for excited states.

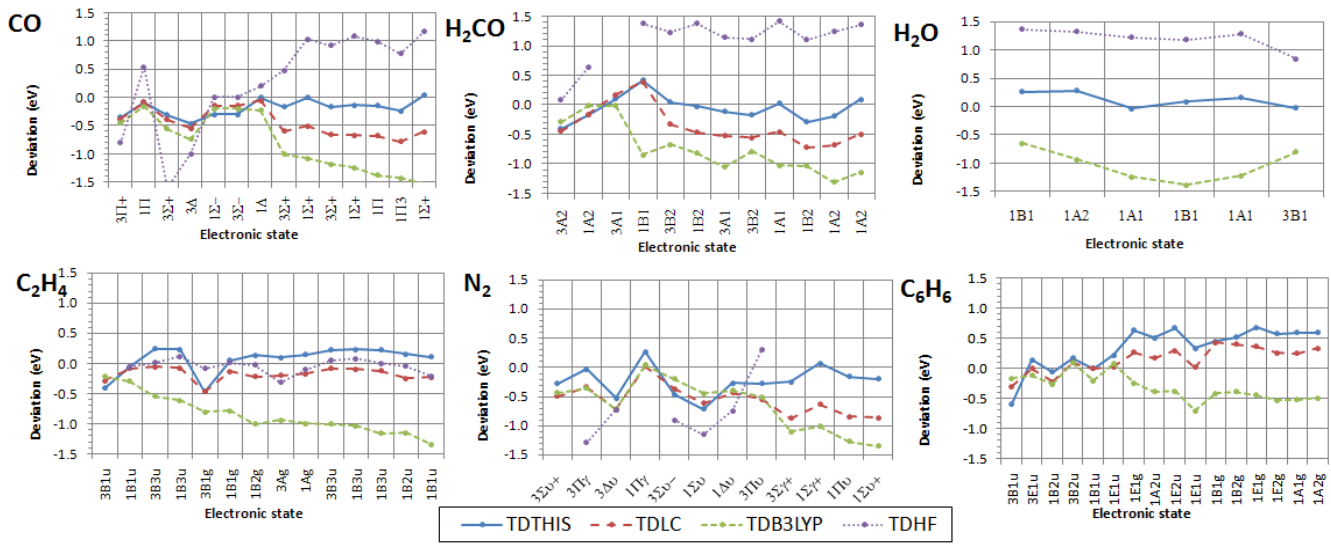

FIGURE 4: THE ENERGY DEVIATION OF CALCULATED VERTICAL EXCITATIONS FROM EXPERIMENTAL EXCITATIONS (THE LATTER TAKEN FROM REF. [74]) FOR SEVERAL MOLECULES.

## C. Inter-molecular Charge Transfer Excitation

Spurious self interaction can impair the description of charge-transfer interaction. The benchmark test for such a problem is the model of Dreuw et al[26], where $C_2H_4$ and $C_2F_4$ molecules considered at a large distance apart $R$. One of the excited states involves transfer of charge from $C_2F_4$ to $C_2H_4$ and this state is of interest. The finite basis set in the model calculation artificially stabilizes the anionic state of $C_2H_4$, so it serves as an excellent model for charge transfer benchmarks. The inter-molecular charge transfer excitation energy must depend asymptotically on the distance between the two molecules as[26]:

$$\lim_{R \to \infty} \Delta E(R) = EA(C_2H_4) + IP(C_2F_4) - \frac{1}{R} \quad (17)$$

Indeed, as reported[26] TDB3LYP could not reconstruct this behavior, as seen in Figure 5, where the TDB3LYP excitation is too low by many electron-volts and clearly deviates from a straight line in the large-R limit. On the other hand, the results of This functional follow the straight line very closely and the infinite R limit yields an excitation energy of 13.1 eV, which is close to the correct result: in the same basis we calculated the EA and IP of the two molecules, obtaining the results shown in Table 7. The value if $EA+IP$ around 13eV is completely consistent with the excited state data.

TABLE 7: THE CALCULATED ELECTRON AFFINITY AND IONIZATION POTENTIAL USING THIS FUNCTIONAL. ALL ENERGIES ARE IN EV.

|  | IP($C_2F_4$) | EA($C_2H_4$) | IP+EA |
|---|---|---|---|
| $-\varepsilon^{HOMO}$ | 9.7 | 3.2 | 12.9 |
| $\Delta SCF$ | 9.7 | 3.4 | 13.1 |

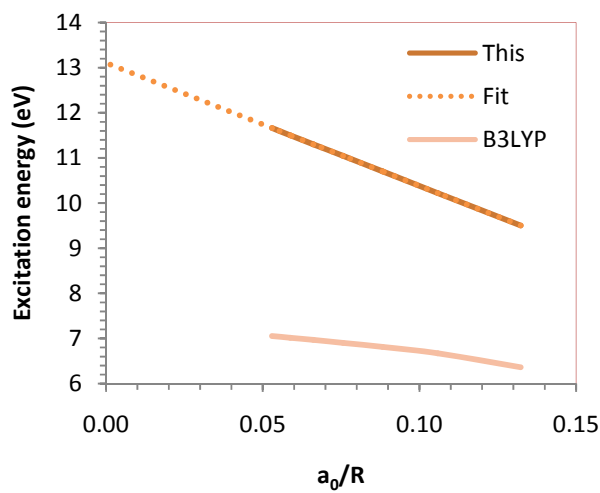

**FIGURE 5: THE EXCITATION ENERGY VS. 1/R IN THE CHARGE-TRANSFER EXCITATION OF $C_2H_4 \cdots C_2F_4$. THE B3LYP RESULT IS TOO LOW AND HAS THE WRONG R-DEPENDANCE.**

## IV. Summary and Discussion

We have developed in this paper a semi-empirical approach to density functional theory which makes use of an explicit exchange term (Eq.(2)) The functional has two empirical parameters and we used the G2(1) database to optimize their values. Then we tested the resulting functional in a variety of other settings, showing that it is balanced and robust across many types of molecular properties. Such a functional does a reasonable job at describing the chemical bond (although it is surely not the best choice for ground-state chemical dynamics), it is capable of describing a variety of electronic properties and excited electronic states. We believe that such a balanced approach to density functional theory is possibly very useful for a variety of fields, such as photochemistry and biochemistry, molecular electronics, materials science.

One point of formal concern is the fact that This functional is not exactly consistent with the homogeneous electron gas, as other functionals are[34-40]. Our initial approach – that of adjusting $\gamma$ was indeed designed to have this exact property conserved, without giving up the correct long range behavior of the potential. However, we found that in practice that theory did not give sufficiently good results. The reason is that we do not have a good approximation for the functional $E_{XC}^{\gamma}$ in non-homogeneous densities (see Eq. (3)). The optimal way to proceed then is to try and improve $E_{XC}^{\gamma}$ beyond the homogeneous case, perhaps by introducing a proper gradient approximation. Since we have a practical way of determining $\gamma[n]$ once $E_{XC}^{\gamma}$ is given, this may be a viable way to proceed. In the current work, we made drastic approximations that allowed us to bypass this problem. The generally encouraging results we obtain, allow us to be optimistic that this type of approach may be the key for better functional development in the future.

**Acknowledgments:** The authors gratefully thank Dr. Yihan Shao for his invaluable assistance and guidance in coding this functional into Q-CHEM 2[43]. We also thank Prof. Martin Head-Gordon for his support and finally Prof. Daniel Neuhauser for his input and in-depth discussions on these issues. We acknowledge support from the Israel Science Foundation.